# CRYPTO MULTI TENANT: AN ENVIRONMENT OF SECURE COMPUTING USING CLOUD SQL

Parul Kashyap and Rahul Singh

Department of Computer Science & Engineering, S.R.M.U., Uttar Pradesh, India

## ABSTRACT

*Today's most modern research area of computing is cloud computing due to its ability to diminish the costs associated with virtualization, high availability, dynamic resource pools and increases the efficiency of computing. But still it contains some drawbacks such as privacy, security, etc. This paper is thoroughly focused on the security of data of multi tenant model obtains from the virtualization feature of cloud computing. We use AES-128 bit algorithm and cloud SQL to protect sensitive data before storing in the cloud. When the authorized customer arises for usage of data, then data firstly decrypted after that provides to the customer. Multi tenant infrastructure is supported by Google, which prefers pushing of contents in short iteration cycle. As the customer is distributed and their demands can arise anywhere, anytime so data can't store at particular site it must be available different sites also. For this faster accessing by different users from different places Google is the best one. To get high reliability and availability data is stored in encrypted before storing in database and updated every time after usage. It is very easy to use without requiring any software. This authenticate user can recover their encrypted and decrypted data, afford efficient and data storage security in the cloud.*

## 1. INTRODUCTION

In this modern period internet is working as a conventional hosting system which is accessible through different services with limited usage and storage. But the current drift in business requires vitality in computing and storage, it causes the development of cloud computing. For the issues of computing, storage, and software, cloud computing proposes new models. This model provides expansion in the environment, allocation and reallocation of assets when desired, storage and networking ability virtually. It satisfies on demand need of the customers. Cloud computing work as a combination of computational paradigm and distribution architecture. The major aim is to provide services like net computing services, flexible data storage containing required resources visualized as a service and delivered over the internet [1] [2]. Cloud computing enhances many features like scalability, availability, adapt progress of demand, teamwork, pick up the pace of development, provide potential for outlay fall through efficient and optimized computing [3][ 4].

One of the important characteristics of cloud computing is multitenancy. This feature is similar in nature with other multiple families on the same platform. It contained large no. Of resources and data of different users which get differentiate by their unique identification. In this multitenancy model certain level of control is provides for customizing and tailoring of hardware and software for fulfilling customer demands. In multi tenancy model physical server partition occur with virtualization. This virtualization feature contains good capability of separation. But still some security issues arise those are data isolation, architecture expansion, configuration self definition and performance customization. This feature favours users to copy, create, drift and roll back virtual machines for running many applications [5] [6]. Like physical machine security, virtual machine security is also important Virtual machine because error in any virtual machine may cause in error in other [7]. Virtual machine as two boundaries- physical and virtual as physical servers [8]. This paper is basically focused on data isolation.

From the aspect of quality of service data security is important (cong et al., 2009). When the customer arises to satisfy their demand, it firstly assures that their data will be secure. Due to which we use encryption algorithm to secure data storage and access. For this AES (advanced encryption standard) 128 bit encryption algorithm is used to encrypt data before storing data in a database. When the customer came to use resources or data on the physical server if it is authorized then firstly decrypted and after that it provides to the user. Multi tenant infrastructure is supported by many Applications, Google is one of them. Google pushes the contents in short iteration. There are large of functionalities are added weekly basis, due to which required update also done weekly basis in a Google cloud (http://www.Google.com/Apps/intl/en/business/cloud.html). When the new features are introduced, they automatically reflected in the browser. As Google sustain by cloud computing, it gets updated to fulfil millions of customer demands. The customers' needs are not only maintained on single site, but other secured centres also, so that when one site fails to fulfil the customer demand it get fulfil by another site. Google supports parallel and fast access from various places. For flexibility and reliability data are stored in various data centres. It is easy to use Google cloud SQL. It does not require any type of software. Google cloud SQL concerns, my SQL instances, which is similar to my SQL. It contains all features as my SQL and other contained features are:

Instances limited up to 10GB, synchronous duplication, importation database, export database, command line tools, highly obtainable, completely managed, SQL prompt.

In this study, we suggest a way of implementing the AES algorithm on the data before storing in the database by the cloud service provider (CSP). Every cloud service seeker should confirm the security criteria with cloud security provider before hosting data in the database. As many tenants cause many changes to their data after every regular interval, so there are many challenges to overcome.

## 2. CHALLENGES TO THE DATA SECURITY IN THE MULTI TENANT MODEL

Cloud basically offers classy storage and admittance climate, but this is not hundred percentages trustworthy; the dispute exists in ensuring the authoritative admittance. As the data in multiple tenant model shared between large numbers of tenants so its storage is an important issue. Each tenant ensures firstly before storing data in the cloud that his data is at a secure place or not. In the current period, few vendor cloud data centres has breached of cloud data security, recent example of it, is the hacking of the nearly 450,000 passwords from the yahoo service called 'yahoo voice'. The exposure surroundings the event states that the primary technique used by the hacker is SQL injection to get the information out of the database. As we know that network medium is the main source of communication, so there is the possibility of sharing of network component among tenants due to resource pooling [9] [10]. Security is the principal unease on multi tenancy model, while best benefits of cloud delivery model is obtained through multi tenancy, where resources is shared by many users demanding of efficiency. SAAS users totally depend on provider for proper security then data security is the major issue [10] [12].

When the big organization comes to store their data, they already get nervous when they hear their data going to store with their competitive organizations. Cloud provider ease their stress by providing multitenant architecture which basically deals with the security issues related to the tenants data from each other. In SAAS or organizational data is serving in plain text and then stored in the cloud. So SAAS provider is one dependable for security of data at the time of serving and storing [13]. There are basically two issues of data security are as follows:

A. Security of data at transmit state.
B. Security of data at rest state.
C. Security of data in transmit state is basically concerned with isolation at network level.
 While security of data at rest state is further divided in four parts where data is residing and security issues can arise.
A. Application level security.
B. User level security, concerned with mobility and replication of data of users.
C. Virtual level security, concerned with unauthorized exposure, unapproved migration, physical compromise.
D. Tenant level security, concerned with the data between the tenants.

## 3. PROPOSED WORK

Multi tenancy is the model where the physical server provides with partition by virtualization. This partitioned server refers as a virtual machine (VM) and the users become tenants. Basically data of tenants are stored in the database, so design of the database is also plays an important role. Typically three ways are available to store data:

- Separate database- each tenant refers separate database.
- Separate schema- each tenant provides a separate logical unit called schema.
- Separate rows- each tenant allocates same database and schema, but each tenant information get separate with their primary key as row wise in the database.

 This paper basically focuses on the third concept of separate rows. In the multi tenant model tenant's data isolation is a great issue. So from the security point view separate database is provided to each of the tenants, but it is less reliable and also time taking. To overcome this penalty a single database is used by all the tenants on the same physical machine. But still there are problems of security of data of tenants from each other which get reduced by providing the concept of cryptography. Through cryptography concept we use AES (advanced encryption standard) algorithm to encrypt data of the tenant before storing in the database. Each row of database separates by their differentiating by their ids.

Each time when the customer comes to store their data, it firstly faces the encryption boundary of AES algorithm provided by the cloud service provider (CSP). Where it gets encrypted and after that it store in database.

## 4. BASE METHODOLOGY

AES algorithm- basically AES is a symmetric block algorithm. This means encryption and decryption is done by the same key. This algorithm can accept block size of 128 bits and use three keys of choice, i.e. 128, 192, 256 bits Based upon which type of version is used by the customer; hence the standard is named as AES-128, AES-192, and AES-256 respectively. The encryption process of AES consists of 10 rounds of processing for 128 bits. All rounds are identical except the last one. AES algorithm begins with a stage of adding round key followed by 9 rounds of four stages and tenth round of three stages. The required stages are as follows:

1. Sub bytes- In this step of sub bytes a look up table is used in which each byte is replaced with.

2. Shift rows- In this step of shift rows, shifting of rows occurs cyclically with a particular offset, while leaving the first unchanged.

3. Mix column- In this step of mixing operations occur using an invertible linear conversion in order to merge the four bytes in each column.

4. Add round key- This step includes derivation of round keys from Rijindael's key agenda and this round key gets added to each byte of the state.

In these last steps are performed up to fourth round, fifth round contains above all step except mix column step as shown in figure 1.

## 4. EXPERIMENTAL METHODOLOGY

There are following steps which require implementing the AES algorithm in cloud to create Google Application:

Step 1: firstly enter the URL http://accounts.Google.com/, then Google user name and password.
Step 2: Select Google Application link (my Application).
Step 3: Choose "Create Application" key, grant Application identifier, Application heading and then get on "Create Application" button. At this instant Application is ready.

Now we implement AES algorithm in the Google cloud:

There are subsequent methods to create database, tables in Google cloud SQL

Step 1: Enter https://code.Google.com/apis/console and go for Google Cloud SQL alternative.
Step 2: Select "New Instance" tab from the right upper corner and popup window displayed.
Step 3: Now enter the instance name and correlated an authorative Application which is formed earlier and then click on "Create Instance" button.
Step 4: Go for instance name to visualize its associated properties.
Step 5: From loading database automatically we select "SQL Prompt" tab.

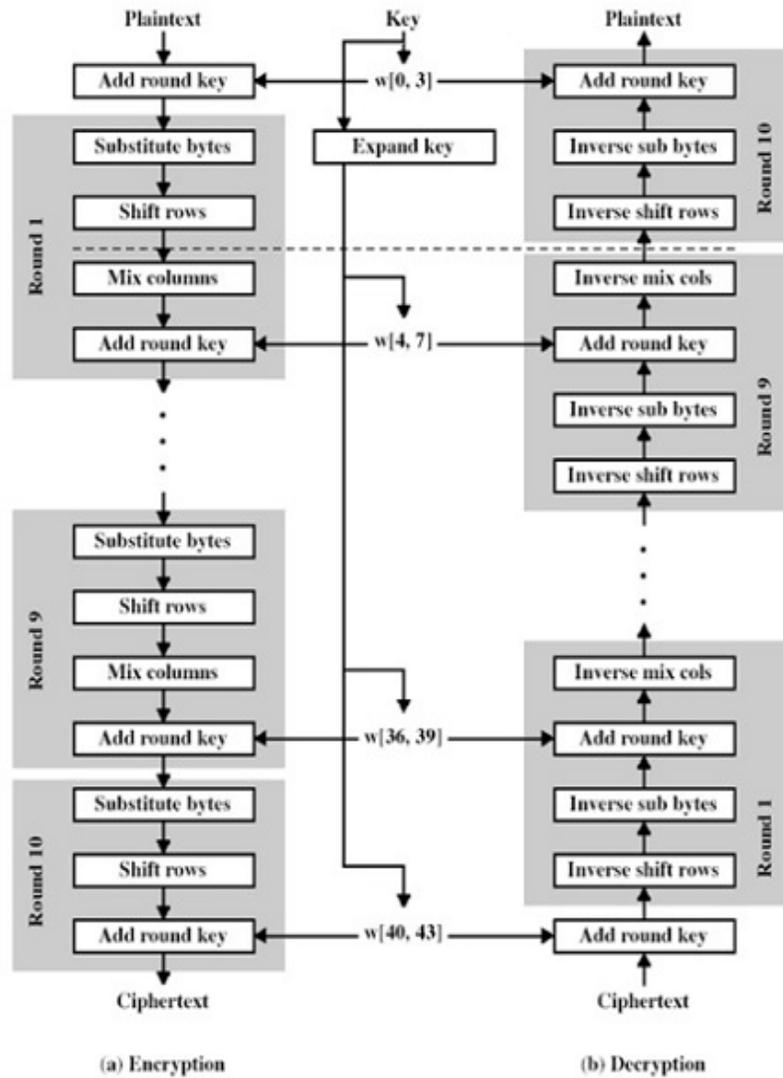

Figure 1.  Encryption and Decryption of AES

Step 6: For creating a database for the Application we use "Create Database" query, all necessary tables are created.

Step 7: Add record to the desired table, we use "Insert Into" query.

Step 8: Generate client interface for the Application.

Step 9: Now write Java code of the AES algorithm to implement algorithm in cloud and debug the Application in Google cloud.

Step 10: Encrypted get sore and decrypted data is displayed while accessing.

# 5. SAMPLE DATA

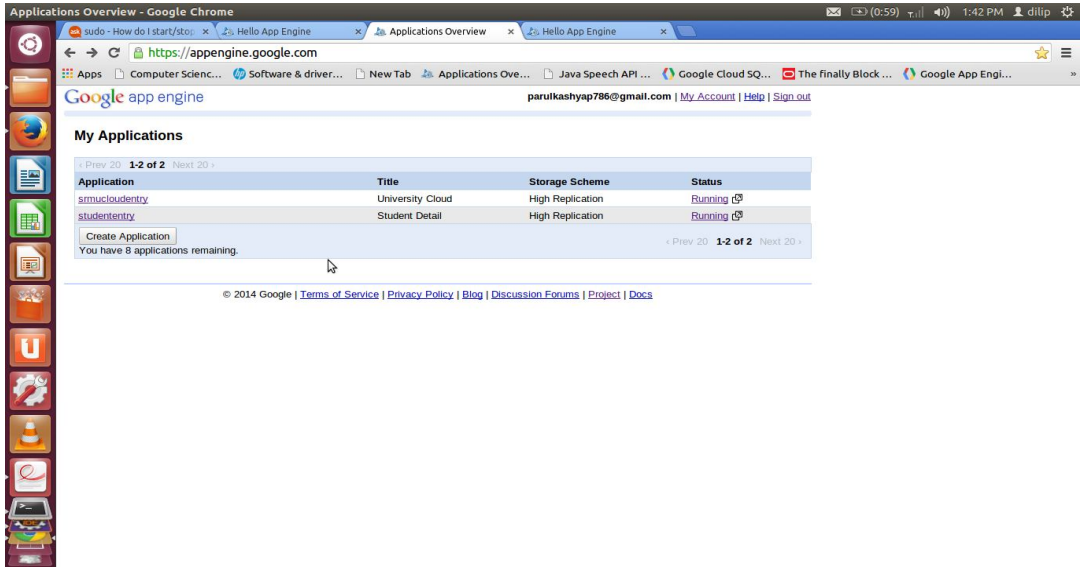

Figure 2. Application "Student Entry" Created in Google App Engine.

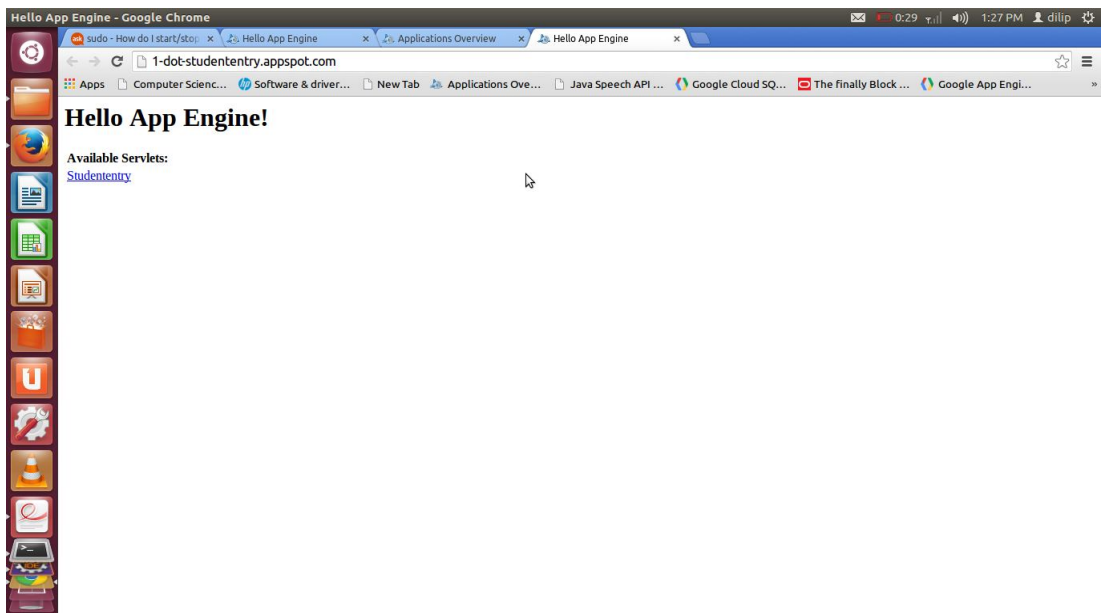

Figure 3. Hello App Engine.

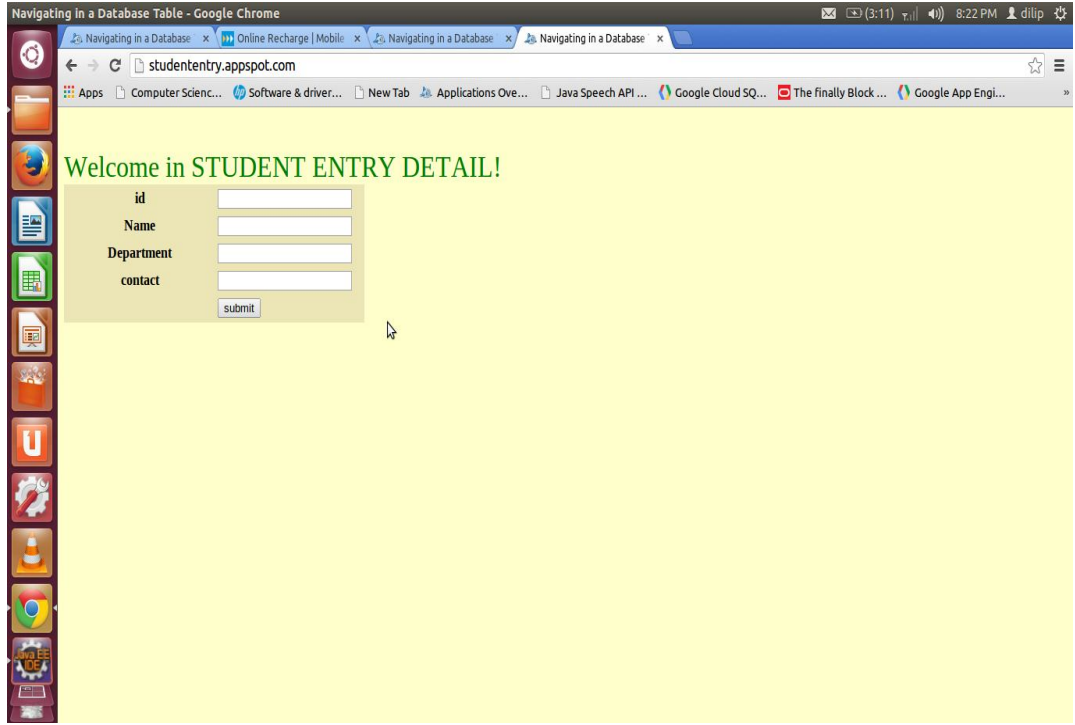

Figure 4. Student Entry Detail.

A User Interface and Application are Created Using Java and Jsp in Eclipse by Accquiring Following Steps are:

Step 1: Database is created in Google cloud named as "Studentry".

Step 2: "Student Entry Detail" table is created in student entry database with its required field like name, id, contact department etc.

Step 3: An application "Student entry" was created in Google app engine by applying above specified steps shown in figure2.

Step 4: To operate the Student Entry Details we designed user interface. User interface uses these details. It firstly choose student link, then Student Entry Detail is displayed as shown in Figure 3 and Figure 4.

Step 5: Now by clicking on "Submit" button the details is received by student class and keys are generated using AES algorithm.

Step 6: Using these generated keys data is encrypted using AES algorithm and then stored in database.

Step 7: At the stage of data retrieval it is decrypted using generated keys.

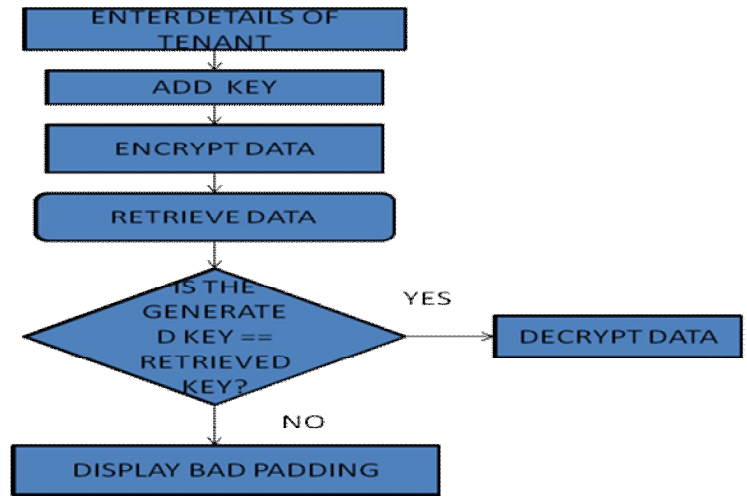

Figure 5.  Execution Flow of Entire Process

## 6. CONCLUSIONS

In our proposed work we mainly focus on data security of tenants so that only the authorized user can access the data.  To provide this feature we use encryption and decryption process of data so that the only legal tenant can access their particular data. Hence, for security related to data of tenants we implement AES (advanced encryption standard) using cloud SQL. The result obtained from experimental methodology proved that AES gives protection for the data stored in the cloud. We make use of AES algorithm and Google App Engine to supply secured data storage, efficiency, assure availability in the condition of cloud denial-of-service attacks and data security in the cloud. This Approach is basically implemented by tenants who are going to store their data in the cloud. This approach is implemented by tenant itself at the cloud security provider (CSP) who stores data in the database.

## ACKNOWLEDGEMENTS

This research paper is made possible through the help and support from everyone, including: parents, teachers, family, friends, and in essence, all sentient beings.

## Authors

**Parul Kashyap** is student of M.Tech., Dept. of  Computer Science and Engg., S.R.M.U., Lucknow, Uttar Pradesh, India and she has completed B.Tech.(Computer Science & Engineering) from RIET Kanpur, India in 2012. Currently, She is working research on Data Security. She has contributed in various research papers in various National and International journals.

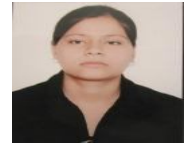

**Rahul Singh**, Asst. Professor, Dept. of Computer Science and Engineering, SRMU, Lucknow, Uttar Pradesh, India. He has completed M.Tech (Computer Science & Engineering) from Motilal Nehru National Institute of Technology, Allahabad, India and B.Tech (Computer Science & Engineering) from Ajay Kumar Garg Engineering College, Ghaziabad, India 2013 and 2010 respectively. His research interests include Cloud computing, Distributed Computing and Mobile computing.

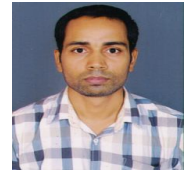